# New hybrid distributed voting algorithm

Nina Evseenko
Moscow State Industrial University
P2PTech
nina@p2ptech.org

*Abstract*—**Increasing data volumes requires additional rating techniques. Reputation systems are the subject of much research. There are various techniques to rate content that facilitate the search of quality content. Page rank, citation index and votes from users are some rating examples. In the article I focus on decentralized vote systems. The article reviews several distributed vote designs. I list the distributed vote requirements. A new hybrid algorithm is proposed which operates in the structured overlay P2P DHT Kademlia network.**

## I. INTRODUCTION

Researching and practical application of distributed systems allows to reduce resource consumption and to deploy large scale and reliable services [1]. An absence of centralization enables load sharing and eliminates a central point of failure.

At the present time, P2P file sharing remains a very popular method of exchanging data among users [2]. Reputation systems are of significant practical interest to marketers [3]. Also, peer-to-peer data exchanging, which is distributed with robust votes, is a very useful tool in view of the current conditions of increasing information content.

According to the statement from [4] users of P2P file sharing systems prefer distributed votes to centralized ones. Moreover, distributed vote systems are the solution of a problem considered in [15], "limited distribution of feedback limits its effectiveness". In view of all of the features of P2P systems, the next section formulates requirements which should be considered to design distributed vote systems for P2P file sharing systems.

The third section is an outline overview of two of the most popular, usable, and almost implemented designs of distributed votes for P2P file sharing systems. In addition to their advantages, I also emphasize their weak sides and suggest solutions that may solve most of the problems.

The following sections describe the design of a new hybrid distributed vote algorithm. The last section describes outstanding works within the suggested algorithm.

## II. REQUIREMENTS TO DISTRIBUTED VOTE SYSTEMS

When designing distributed vote systems the following requirements should be considered:

- Robustness to misbehaving users, both selfish and malicious [4]
- Solving problem with free-riders [5].
- Robustness to network dynamics (churn) [6]
- Incentive to vote
- Scalability
- Security
- Low local resource consumption
- Low traffic consumption
- Take into consideration users behind NAT
- Up-to-date vote information

## III. OVERVIEW OF EXISTING DESIGNS

Voting as a phenomenon, decentralized and centralized, is the subject of much research. The work [7] describes taxonomy of vote systems. An architecture of a vote system depends on the type of users, the type of network being utilized (for example, dealing with unstructured networks involves additional techniques that eliminate the free-rider problem), the purpose of the votes. Hereinafter, I will review distributed vote systems in the production stage designed for P2P file sharing systems.

### A. Robust vote sampling in P2P media distribution systems

The article [8] describes a distributed vote system designed for Tribler which is the BitTorrent media client [9].

Voting is realized with a two-level scheme. First, users vote for moderators; then users receive votes from selected moderators.

The disadvantage of this solution is that manual rating of moderators becomes more difficult for users as the network grows.

As a transport, the design suggests to use randomized gossip to cope with network churn that decreases the precision of votes and increases the time to obtain information.

Despite using signed communication channels between peers the vote system is subject to Sybil attack.

### B. uTorrent's vote system

A specification of this vote system has not been published yet. During the analysis made within BTDigg research project, the following features of the vote system were extracted:

- The voting system utilizes the BitTorrent DHT network
- A voting packet links an info-hash and a user's vote

- Votes are 5-star (users cannot express negative opinion concerning spam content).
- Votes are announced periodically
- A vote can be set once

In the estimation about 1% of BitTorrent DHT packets are vote packets (Fig. 1). From a user's point of view the main disadvantage of this vote system is inability to detect spam using received votes.

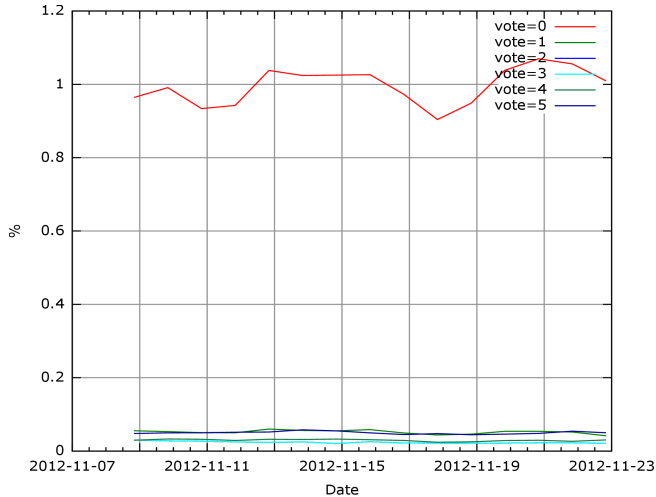

*Figure 1.    Percentage of vote packets in BitTorrent DHT traffic*

## IV. DESIGN OF A NEW HYBRID DISTRIBUTED VOTING ALGORITHM

The algorithm was named hybrid because it inherits the current techniques and eliminates some their disadvantages. Implementation was based on structured P2P DHT Kademlia [10] network according to the BitTorrent specification (Mainline) [11]. The following commands were added:

- announce_vote — the command to announce a vote
- get_votes — the command to get votes

The advantage of utilizing of the BitTorrent DHT network is that I can eliminate free-rider problem from consideration because nodes have an equal responsibility in the network and because the network is robust to high churn rates [12].

To be resilient to malice and colluding I use hashed document ID as a DHT key which is an info-hash in the case of deploying in the DHT BitTorrent network. Hence, a node doesn't know what kind of votes, or for which document, it is storing. SHA1 might be used as a hash function [13].

There are various types of ratings: star, multilevel, continued and others. For the hybrid vote system, I selected the negative/positive continued rating model. In this case a user may express either positive (like) or negative opinion (unlike). This model can be easily expanded to multilevel continued rating model. Furthermore it is similar to the widely used model of "likes" in Facebook, so it is the familiar method of voting for a majority of users.

Votes are stored separately in two probabilistic counters. One counter collects negative votes, another - positive votes. The counters are built based on HyperLogLog algorithm [14] which consumes very few resources and provides optimal accuracy. The hybrid algorithm considers one vote per an IP address in condition that there are not too many users behind the same NAT and sharing the same IP address.

A node (user) stores its votes permanently and periodically announces to its k-nearest neighbors in the DHT network. A user can rate any content at any moment linking his/her vote with the info-hash, but only once. In addition to its own votes, a node stores the votes of other participants announcing their votes to the node. When a user needs to know the rating for content, he/she evaluates hashed info-hash and selects the k-nearest nodes which are closest to the calculated hash value. The user sends get_votes command to each of the evaluated neighbour nodes. When every k-node has responded or timed out, the user aggregates votes and applies filter procedures to eliminate spam.

Details.

### 1) Data structure for storing votes

Fig. 2 illustrates the data structure for storing information about votes. When the announce_vote command is received, a node selects or creates the corresponding hashed info-hash element in its list and the related ring buffer. Then the block matching a current hour is selected from the proper ring buffer. If a difference between a current time and time when the block was created is more than 24 hours, then the counters in the block are reset. In such a way, votes remain up-to-date. If the nodes announce their votes permanently and periodically, the votes will remain current.

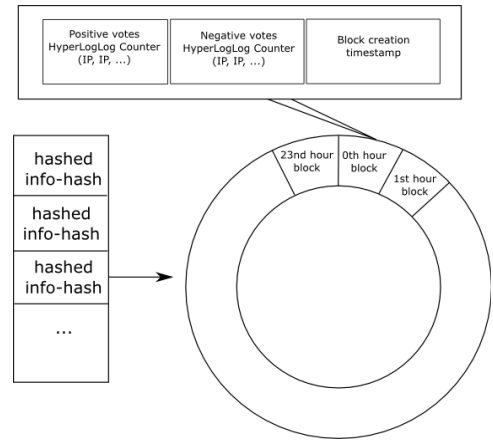

*Figure 2.    Data structure for storing votes*

### 2) Replication

The evaluation of k-nearest neighbors is done using the same method as described in the BitTorrent DHT specification

(the nearest node IDs to a hashed info-hash). Such a replication solves the problem with network dynamics and provides a method to detect malicious nodes sending wry data.

*3) Getting votes*

To get votes, a node selects k-nearest neighbors and sends the get_votes command to them. Votes are collected by a user's request. However, the replication constant k must be optimal and each of k-nearest nodes must be storing up-to-date information about announced votes.

## V. FURTHER WORK

This article is an introduction to suggested methods. Further investigations will be done concerning the joining nodes behind NAT with same IP address, aggregation of votes, a method of detection of malicious nodes, period of announcing, replication level, analysis of accuracy and resource consumption.